*Economic effects on households of an augmentation of the cash back duration of real estate loan*

-

*Effets économiques sur les ménages d'une augmentation des durées maximum d'emprunts immobiliers (entre 30 à 80 annuités)* – **Working papers**

Hugo Spring-Ragain

CEDS

# Abstract:

This article examines the economic effects of an increase in the duration of home loans on households, focusing on the French real estate market. It highlights trends in the property market, existing loan systems in other countries (such as bullet loans in Sweden and Japanese home loans), the current state of the property market in France, the potential effects of an increase in the amortization period of home loans, and the financial implications for households.

The article points out that increasing the repayment period on home loans could reduce the amount of monthly instalments to be repaid, thereby facilitating access to credit for the most modest households. However, this measure also raises concerns about overall credit costs, financial stability and the impact on property prices. In addition, it highlights the differences between existing lending systems in other countries, such as the bullet loan in Sweden and Japanese home loans, and the current characteristics of home loans in France, notably interest rates and house price trends.
The article proposes a model of the potential effects of an increase in the amortization period of home loans on housing demand, housing supply, property prices and the associated financial risks.

In conclusion, the article highlights the crucial importance of household debt for individual and economic financial stability. It highlights the distortion between supply and demand for home loans as amortization periods increase, and the significant rise in overall loan costs for households. It also underlines the need to address structural issues such as the sustainable reduction in interest rates, the stabilization of banks' equity capital and the development of a regulatory framework for intergenerational lending to ensure a properly functioning market.

# Résumé :

Cet article examine les effets économiques d'une augmentation de la durée des prêts immobiliers sur les ménages, en se concentrant sur le marché immobilier français. Il met en évidence les tendances du marché immobilier, les systèmes de prêt existants dans d'autres pays (tels que les prêts in fine en Suède et les prêts immobiliers japonais), l'état actuel du marché immobilier en France, les effets potentiels d'une augmentation de la durée d'amortissement des prêts immobiliers, ainsi que les implications financières pour les ménages.

L'article souligne que l'augmentation de la durée de remboursement des prêts immobiliers pourrait réduire le montant des mensualités à rembourser chaque mois, facilitant ainsi l'accès au crédit pour les ménages les plus modestes. Cependant, cette mesure soulève également des inquiétudes concernant les coût globaux du crédit, la stabilité financière et l'impact sur les prix de l'immobilier. De plus, il met en évidence les différences entre les systèmes de prêt existants dans d'autres pays, tels que le prêt in fine en Suède et les prêts immobiliers japonais, et les caractéristiques actuelles des prêts immobiliers en France, notamment les taux d'intérêt et l'évolution des prix de l'immobilier.
L'article propose une modélisation des effets potentiels d'une augmentation de la durée d'amortissement des prêts immobiliers sur la demande de logements, l'offre de logements, les prix de l'immobilier et les risques financiers associés.

En conclusion, l'article souligne l'importance cruciale de l'endettement des ménages pour la stabilité financière individuelle et économique. Il met en évidence une distorsion entre l'offre et la demande de prêts immobiliers avec l'augmentation de la durée d'amortissement, ainsi que l'augmentation significative du coût global du prêt pour les ménages. Il souligne également la nécessité de prendre en compte les problèmes structurels tels que la diminution durable des taux d'intérêt, la stabilisation des fonds propres des banques et l'élaboration d'un cadre réglementaire pour la passation de crédits intergénérationnelle afin de garantir un marché fonctionnant correctement.

The French real estate market is a key economic sector, accounting for around 10% of GDP and having a major impact on the daily lives of households. For several years now, property prices have been rising steadily in major cities, making home ownership increasingly difficult for modest and middle-income households. Against this backdrop, one way of facilitating access to home ownership is to increase the number of years of amortization on real estate loans, in order to give households more leverage to invest.

This measure, which consists in extending the repayment period of home loans, would reduce the amount of monthly repayments to be made, thus facilitating access to credit for the most modest households. However, this measure also raises important questions in terms of household indebtedness, financial stability and the impact on property prices. In addition, we need to consider the impact of interest rates, which have returned to a high level, on the overall cost of the loan for households.

We must therefore try to understand the various interactions of mortgage annuities on supply and demand, and not only on the risk inherent in long-term repayments.

## I. Overview of existing systems

In many Anglo-Saxon countries, a new type of loan - the bullet loan - is becoming increasingly widespread. In fine loans are a form of property loan in which the borrowed capital is repaid in a single instalment at the end of the loan term, rather than in regular repayments as in the case of an amortizing loan. With a bullet loan, the borrower repays only the interest during the term of the loan, resulting in lower monthly payments than with an amortizing loan.

In fine loans are generally used to finance rental investments, as they enable borrowing interest to be deducted from rental income, thereby reducing the tax base. However, they also present higher risks for the borrower, who must have sufficient capital to repay the loan at maturity. However, if the borrower defaults, the bank may require the sale of the property to repay the loan. Although this type of loan is present in France, it is in very small numbers (less than 5% according to the Banque de France), and is much more prevalent in "anglo-saxon" countries.

- *Swedish loan system*

The Swedish system is a benchmark for bullet loans, and was the talk of the town when the IMF sounded the alarm in 2013 about the average 140-year maturity of home loans in Sweden.

In 2016, the Swedish Parliament decided to restrict the maturity of a loan to 105 years for fear of seeing an unprecedented property bubble burst, with the household debt ratio reaching 177%.

In this case, it would be logical to imagine that Sweden has a very low homeownership rate, in contrast to countries where credit is shorter and therefore less costly in fine. However, Eurostat has established the opposite, and Sweden is one of the top 10 European countries with the highest rate of home ownership. In fact, 63.6% of Swedes (9th place) own their own home, compared with 65.1% in France (8th place).

- *Japonese loan system*

Real estate loans in Japan are characterized by their particularly long amortization periods, which can reach 35 years or more. This practice has been commonplace in the country for several decades, thanks to historically low interest rates (see figure below) and a culture of long-term borrowing.

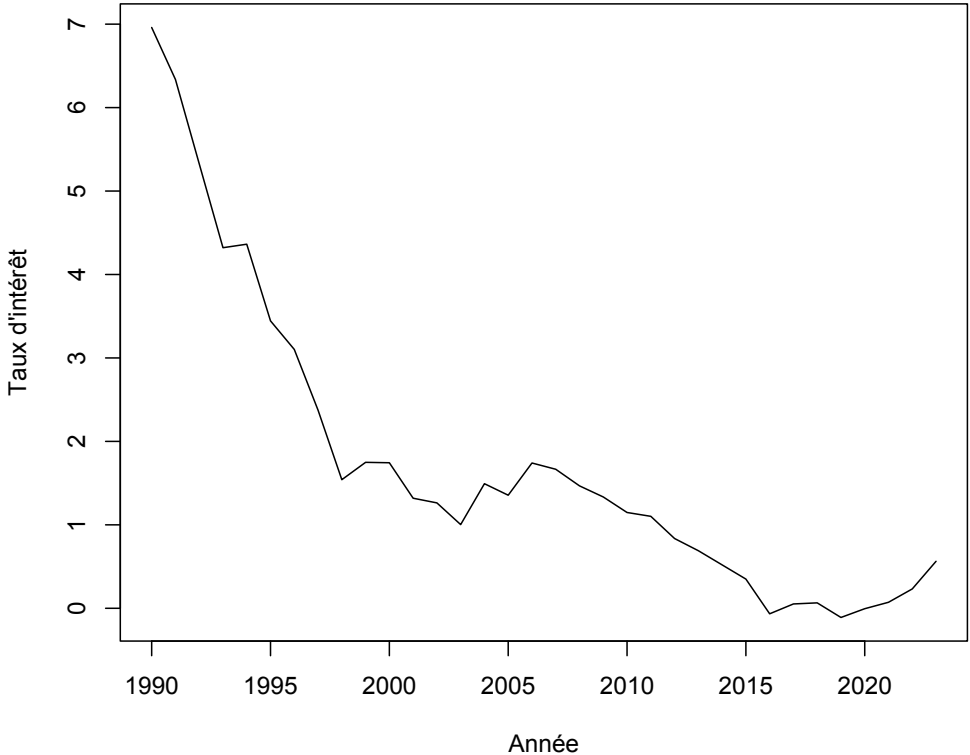

*Figure 1 - Japanese interest rates over the period 1990-2023 - Source: OCDE*

Japanese home loans are generally granted at fixed interest rates, giving borrowers a degree of stability in their monthly repayments. However, interest rates can vary considerably depending on the term of the loan.

Japanese home loans are also characterized by strict collateral requirements. Japanese banks generally require borrowers to provide collateral in the form of a mortgage on the property purchased. Banks may also require additional collateral, such as a personal guarantee or life insurance.

**II. Overview of the French real estate market**
- Current features of home loans in France (interest rates, amortization periods, etc.).

For decades, property loans have been a key financial tool for households in France and around the world, enabling them to buy their own home without having to finance their entire property at once. The French mortgage is not a static tool; it evolves in line with legislation and the economy, particularly interest rates. Here, we'll take a look at the changes that real estate loans have undergone in terms of their various indicators.

*Interest rates*

Over the past decade, the Eurozone, and France in particular, have benefited from particularly low interest rates on home loans, in order to boost property purchases, which were at rock bottom following the subprime crisis.

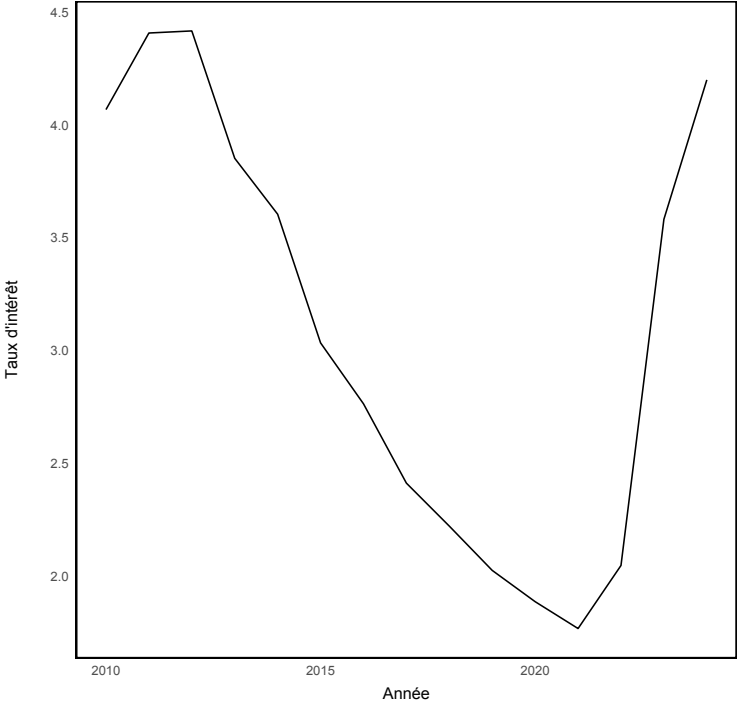

*Figure 2- Home loan interest rates in France 2010-2024 - Source: Banque de France*

Figure 2 highlights what was the golden age of property lending, with the massive influx of loans at less than 2% interest. It's easy to see how post-sovereign debt crisis interest rates fell from 2010 until 2021, when they peak at around 4.2%.

- Property price trends in France in recent years

The figure below shows the evolution of property prices based on an index of 100 in 2015 over the period 2000-2023.

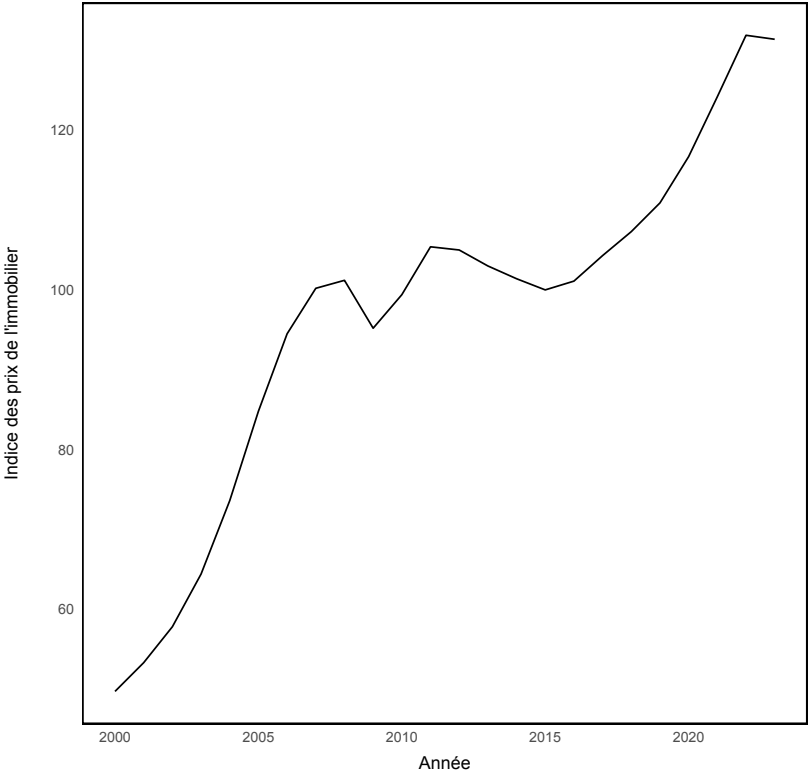

*Figure 3 - Change in property prices based on an index of 100 in 2015 - Sources: INSEE, L'Observatoire Crédit Logement*

Property prices rose sharply between 2000 and 2007, increasing by almost 50% in just seven years. This was followed by a slight downturn in 2008 and 2009, linked to the global financial crisis.

From 2010 onwards, property prices began to rise again, but more moderately than in previous years. This upward trend continued until 2020, when prices accelerated again, rising by more than 10 index points in just two years.

This trend in property prices has major consequences for the real estate market and for households' access to home ownership. Indeed, rising prices can make it increasingly difficult for low-income households or first-time buyers to purchase a property. Coupled with the sharp rise in interest rates seen above, it is economically normal to see a downturn in the property market.

- Modeling current housing supply and demand

D = demand
S = offer

The first step in our reasoning will be to calculate the supply (S) and demand (D) for home loans. Moreover, we start from the initial postulate that $D \neq S$. We will then try to verify this non-equilibrium or, in the opposite case, this equilibrium.

In order to calculate D, it is necessary to define the various variables that are essential to the development of the model:

Average annual household disposable income (Y) = 50 000

Interest rate (r) = 0.035

Average price for a property with an average value of 60 m2 ($P_{prop}$) : 190 680 €

Current amortization period (N) = 20 ans = 240 mois

$$D = f(Y, r, P_{prop}, N)$$

With current indicators, we can estimate demand as follows, integrating the various variables.

$$D = \alpha + \beta_1 * Y + \beta_2 * r + \beta_3 * P_{prop} + \beta_4 * N + c$$

$D = 20000 + 0.001 \cdot 50\,000 + (-160000) \cdot 0.035 + (-0.0025) \cdot 190680 + 3000 \cdot 240 + 0.1$

$D = 20000 + 50 - 5600 - 476.7 + 720000 + 0.1$

$D \approx 733\,973$

Based on the above calculation, we can establish a loan demand of 733 973.

Next, let's integrate the indicators that will be useful for modeling S mortgages:

GDP = 2 779 billions of euros

Property price index (IN) (base 100 en 2015) = 128,9

Inflation rate (INF) = 0.039

$$S = f(r, GDP, IN, INF, D, N)$$

$$S = \alpha + \beta 1 \cdot r + \beta 2 \cdot GDP + \beta 3 \cdot IN + \beta 4 \cdot INF + \beta 5 \cdot D + \beta_6 \cdot N$$

$S = 641.777 + (-50) \cdot 0.035 + 0.0025 \cdot 2{,}779{,}000 + 30 \cdot 128.9 + 100 \cdot 0.039 + 0.01 \cdot 734{,}949 + 22 \cdot 240$

$S = 641.777 + (-1.75) + 6{,}947.5 + 3{,}867 + 3.9 + 7{,}349.49 + 5{,}280$

$S = 24\,077.917$

$S \approx 24{,}077917 \; billions \; of \; euros$

The calculation of supply is, at this stage, limited to a value in billions of euros. We can simply divide this number by the average price of a property in France (190,680 euros) to obtain a theoretical value of property loans that can be provided by banks. However, it's important to note that very few loans include the full price of the property, so we'll subtract 30% from this value, representing the average household's contribution to the loan amount.

So we have :

$P_{prop-30\%} = 133\,476$

$S_{/prêt} = 2{,}4077E + 10 \,/\, 133\,476$
$S_{/prêt} \approx 180\,391$

In this way :

$733\,973 \neq 180\,391$

So, as we assumed in the introduction to this chapter $\boldsymbol{D \neq S}$.

There is therefore a significant distortion between supply (S) and demand (D) for mortgages. But does this hinder the formation of a properly functioning market?

The answer is obviously more complex than a simple yes/no binary. Supply and demand distortion means that the market is driven by supply and, in the case of home loans, by the banks. However, as with any contract, each party must be certain, or almost certain, that the risk linked to supply or demand is under control. This is where the distortion between supply and demand comes into play: households are, for the most part, sure that financing from banks is secure. On the other hand, banks will be more reluctant to finance profiles that involve difficulties or appear to present long-term risks.

**III. The economic effects of an increase in the number of amortization years for home loans**

In order to understand the direct economic effects of longer loan maturities, we need to calculate the economic effects of longer loan maturities. We'll need to calculate the effects on housing demand, housing supply, property prices and related financial risks. This overview will enable us to draw up an initial economic assessment of the effects of a change in loan amortization policy.

We now need to estimate the mathematical impact of this increase by modifying certain variables, in particular the amortization period from 240 to 720 months, in order to show the changes linked to the increase in the repayment period.

Loan amortization period (N) = 60 years = 720 months

- Impact on housing demand

To calculate the evolution of demand as a function of the number of annualities, we'll use the same formula, adapting the different variables and coefficients. We'll then calculate demand for $N_{60}$ and simulate staying with the results over $(N_{21}, N_{22}, N_{23}, N_{24}, \ldots, N_{60})$.

$$D_{N60} = \alpha + \beta_1 * Y + \beta_2 * r + \beta_3 * P_{prop} + \beta_4 * N_{60} + c$$

$Y = 50\,000\ euros$
$r = 0{,}035\ (3{,}5\%)$
$P_{prop} = 190\,680\ euros$
$N_{60} = 720\ months$

$D_{N60} = 24\,000 + 0{,}005 * 50\,000 + (-128\,000) * 0{,}035 + (-0{,}0010) * 190\,680 + 5\,000 * 720 + 0{,}1$

$D_{N60} = 24\,000 + 250 - 4\,480 - 190{,}68 + 3\,600\,000 + 0{,}1$

$D_{N60} \approx 3\,619\,579$

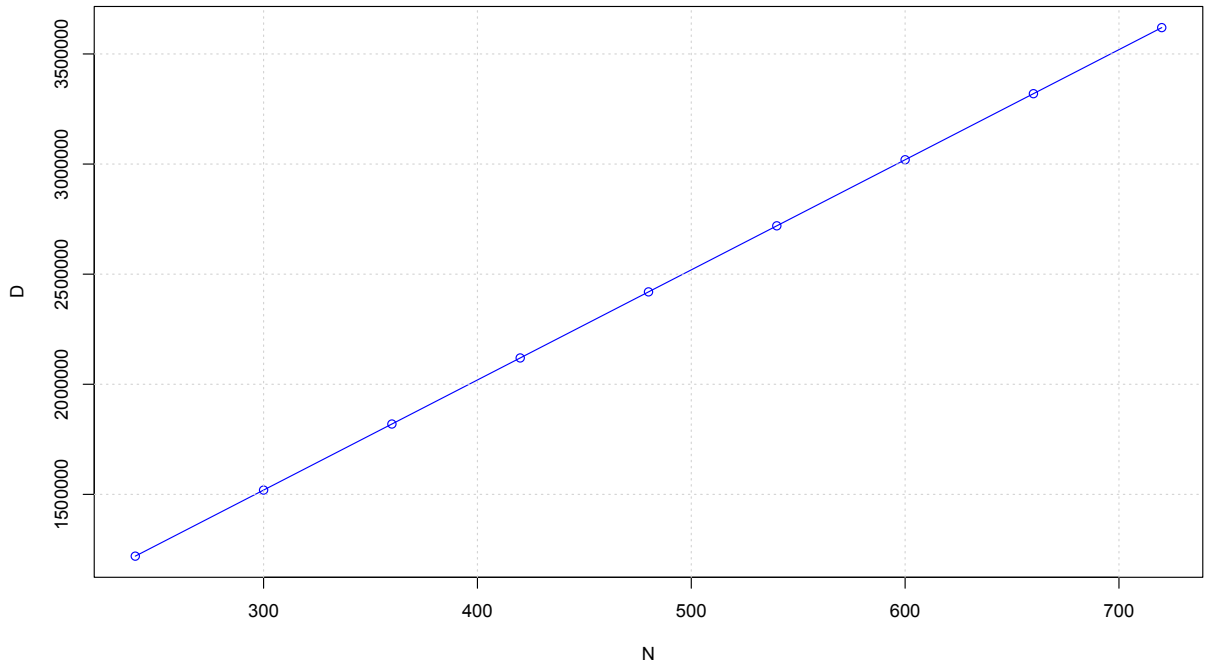

*Figure 4 - Simulation of home loan application with amortization from 20 to 60 years*

Here, it's easy to see that the longer the amortization period, the more likely households are to apply for a mortgage. In fact, as we'll see later, the possibility of spreading the price of a property over a longer period of time drastically reduces household debt levels.

Now that we've simulated consumer demand for mortgages with amortization periods of over 20 years in each case, we need to do the same for the supply of mortgages that banks are prepared to grant.

So, as for D, we calculate $N_{60}$ and simulate $(N_{21}, N_{22}, N_{23}, N_{24}, ..., N_{60})$ while adapting this formula to demand $D_{60}$ and $(D_{21}, D_{22}, D_{23}, D_{24}, ..., D_{60})$ .

$$S_{N60} = \alpha + \beta 1 \cdot r + \beta 2 \cdot GDP + \beta 3 \cdot IN + \beta 4 \cdot INF + \beta 5 \cdot D_{60} + \beta_6 \cdot N_{60}$$

$S_{N60} = 641.777 + (-50) \cdot 0.035 + 0.0025 \cdot 2{,}779{,}000 + 30 \cdot 128.9 + 100 \cdot 0.039 + 0.01 \cdot 2{,}173{,}923 + 22 \cdot 720$

$S_{N60} = 641.777 + (-1.75) + 6{,}947.5 + 3{,}867 + 3.9 + 21{,}739.23 + 15{,}840$

$S_{N60} = 49{,}037.657$

$S_{N60} = 34{,}647657 \; billions \; of \; euros$

*Either*

$S_{N60} = 3,46476E + 10 / 133476$

$S_{N60} = 367\ 389$

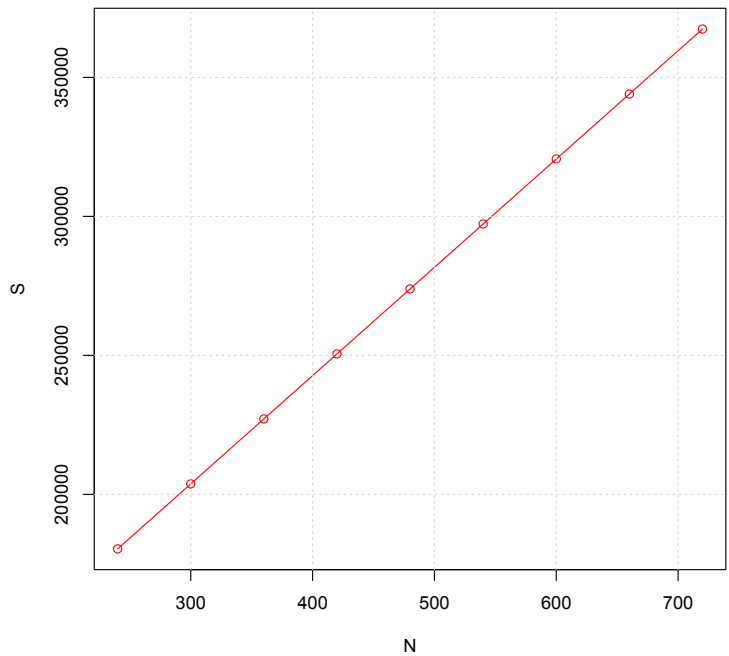

*Figure 5 - Simulation of home loan offer with amortization from 20 to 60 years*

Figure 5 shows the evolution of supply as a function of mortgage amortization years. Although supply is increasing, it remains moderate, with a 43.5% variation from 20 to 60 years' amortization. So adding 40 years of amortization to home loans doesn't even come close to doubling the annual loan supply.

Now we need to study the relationship between supply and demand to see the impact of increasing the amortization period. This analysis will give us an initial idea of the effectiveness of such an economic approach.

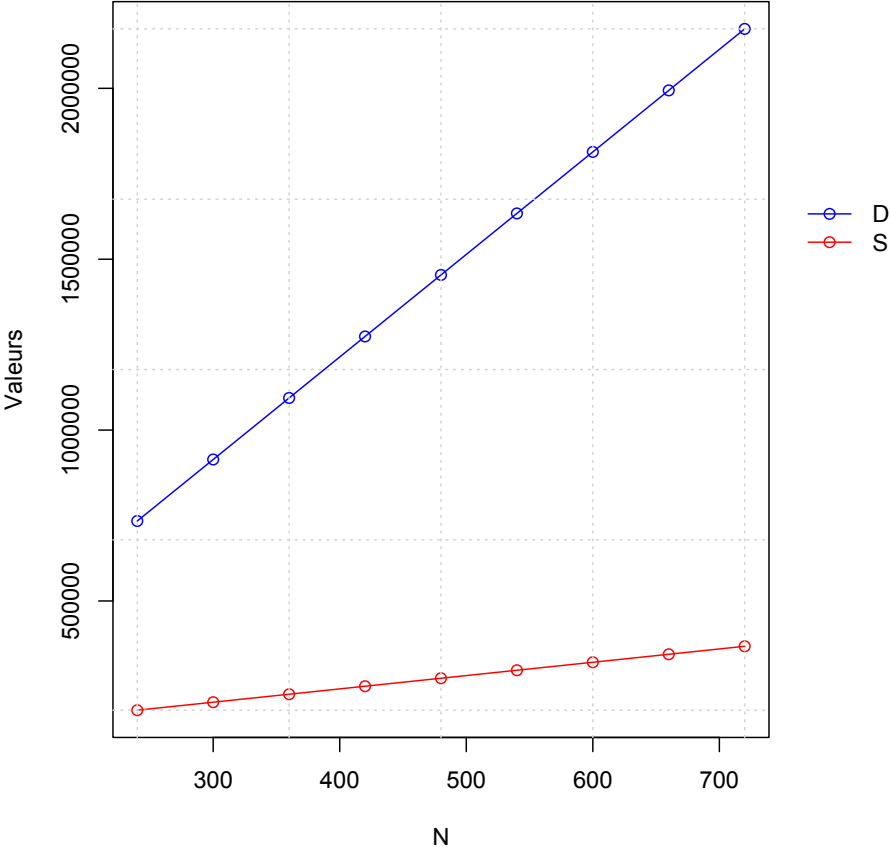

| Years | 20 | 25 | 30 | 35 | 40 |
|---|---|---|---|---|---|
| N | 240 | 300 | 360 | 420 | 480 |
| D | 733 973,40 | 913 973,40 | 1 093 973,40 | 1 273 973,40 | 1 453 973,40 |
| Variation D | | 24,52% | 19,69% | 16,45% | 14,13% |
| S | 180 391,39 | 203 766,60 | 227 141,82 | 250 517,03 | 273 892,24 |
| Variation S | | 12,96% | 11,47% | 10,29% | 9,33% |
| Years | 45 | 50 | 55 | 60 | |
| N | 540 | 600 | 660 | 720 | |
| D | 1 633 973,40 | 1 813 973,40 | 1 993 973,40 | 2 173 973,40 | |
| Variation D | 12,38% | 11,02% | 9,92% | 9,03% | |
| S | 297 267,46 | 320 642,67 | 344 017,88 | 367 389,35 | |
| Variation S | 8,53% | 7,86% | 7,29% | 6,79% | |

*Figure and Table 1 - Comparison of S and N evolutions according to the previous model*

This comparison between supply and demand and their respective evolution highlights a break in the effect that increasing the amortization period could have on paper.

In fact, this comparison reveals an almost inescapable distortion of the market. As the market is already highly unbalanced, as we have seen, this imbalance will become more pronounced with each additional year of amortization proposed.

Secondly, if we are to have a clearer picture of the economic impact of this structural change in real estate lending, we need to consider the impact on real estate prices and on household debt.

- Effects on household debt and financial stability

For this final stage of the demonstration, we'll need to present the mathematical model in its entirety in order to consolidate the analysis. In this way, we will include this variables:

$P_{prop-30\%} = 133\ 476$

$Number\ of\ payments\ per\ year = n = 12\ mois$

In this section, we'll take a look at the total cost of credit, indebtedness and risk for households in the case of an increase in borrowing duration.

First, we'll use the monthly payment formula (M(N)) to calculate the amount to be repaid each month by the borrower:

$$M(N) = \frac{P_{prop-30\%} \cdot \frac{r}{n}}{1 - \left(1 + \frac{r}{n}\right)^{-N}}$$

Following the monthly calculation (M(N)), we now need to calculate total indebtedness (E(N)) of households, providing an overall view of the loan:

$$E(N) = M(N) \cdot N \cdot n$$

To get an overall view of the economic impact of loans on household finances, we'll use the evolution of the total cost of the loan (A) :

$$A = \left(\frac{E - E_{min}}{E_{min}}\right) \cdot 100$$

Now that we know the total cost of the loan for households, we need to calculate the debt ratio ($R_d$) in order to have a reliable indicator of the financial risk that households may be exposed to. :

$$R_d = \frac{M}{\frac{Y}{12}}$$

Calculating the repayment capacity ($C_r$) will then enable us to calculate a reliable composite risk index:

$$C_r = \frac{1}{R_d}$$

Now, taking into account all the above elements, we can model a composite risk index ($I_r$) :

$$I_r = w_1 \cdot R_d + w_2 \cdot A + w_3 \cdot C_r + w_4 \cdot \frac{N}{\max(N)}$$

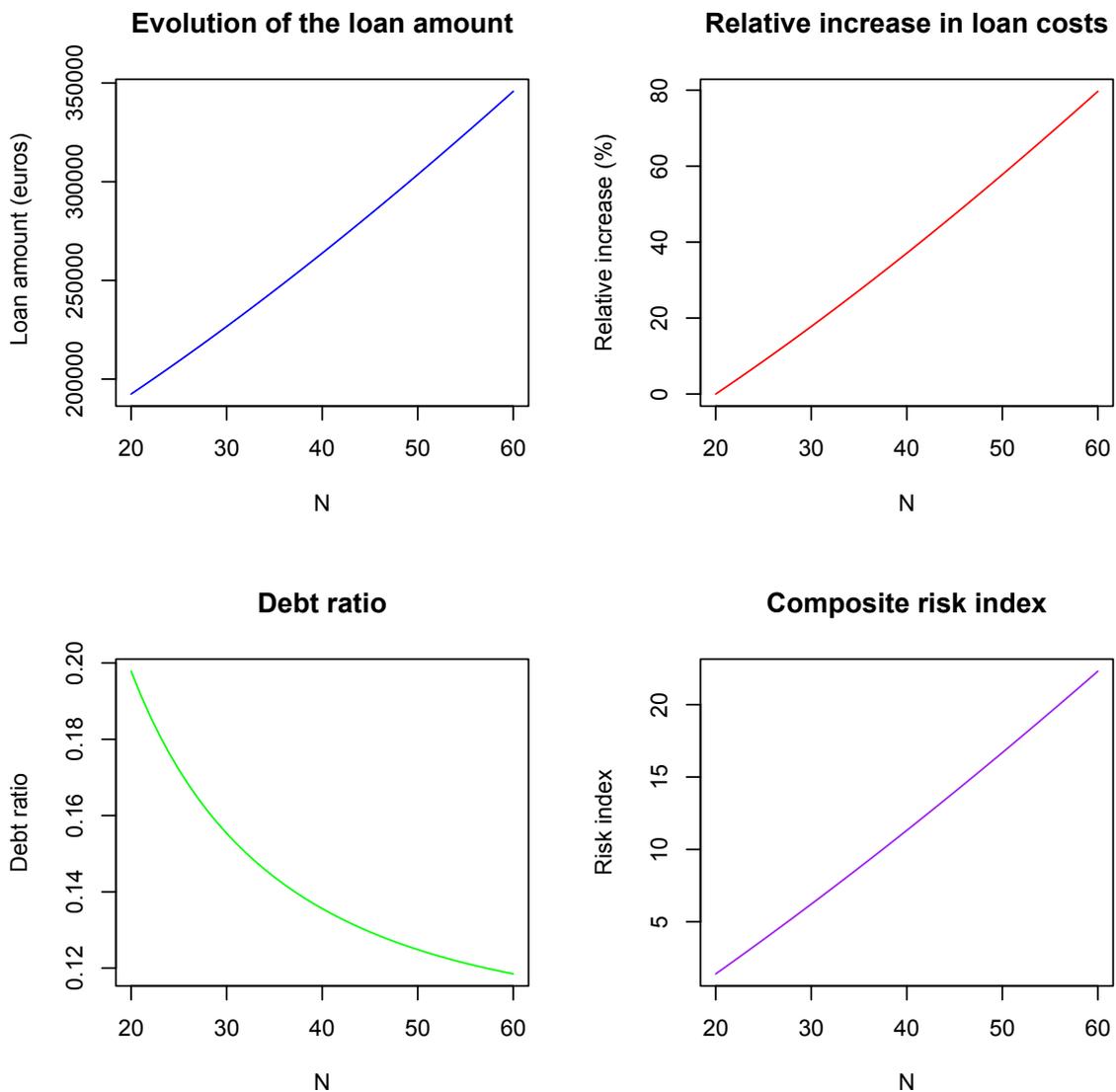

*Figure 6 – Modeling the impact of longer amortization periods on mortgage risks*

Figure 6 provides an overview of the model we have just presented. This model shows several strong paradoxes concerning this increase in the number of years of amortization of home loans.

First, let's look at the cost of the loan as a function of the number of years of amortization. The trend highlighted by the model is an extremely significant increase in the cost of the loan for the household. In fact, by combining capital repayment and interest payments, the cost of the loan will increase by 80% over a 60-year repayment period, compared with a loan repaid over 20 years, as is currently the case.

Secondly, household indebtedness will fall drastically. From a monthly repayment of 801 euros on a 190,680-euro loan spread over 20 years, the monthly repayment is 480 euros on a 60-year loan. It therefore appears that increasing the repayment period for mortgages has the effect of increasing the purchasing power of households, in particular by lowering their property-related expenses.

As for financial risk, it seems clear that it increases progressively with the overall cost of the loan. However, we must not only correlate the increase in risk with the increase in the overall cost of credit, it is also clear that the longer a loan takes to repay, the greater the risk of borrowers defaulting. What's more, repayment at age 60 will, in a number of cases, have a direct impact on the borrower's next generation, thereby also increasing risk.

In addition, the legal issue of taking over the loan will need to be looked into in cases where the borrower has no descendants, or where descendants refuse the property and therefore the loan.

**IV. Conclusion**

Household debt is a crucial issue for individual and economic financial stability. In this article, we have explored various mathematical formulas for modeling the impact of loan amortization periods on the supply and demand for home loans, as well as on total indebtedness, the cost of the loan and the associated risks.

Firstly, we found a significant distortion between supply and demand for home loans, with this distortion increasing as the number of annual installments increases. Supply and demand as calculated therefore seem difficult to reconcile, and there is a real risk that many households will be unable to finance their purchase.

On the other hand, we have seen that the increase in annual installments raises the overall cost of the loan for households to 80%. While this cost seems difficult for households to bear, overall it seems to represent a sharp drop in household debt.

All in all, at current interest rates, it would appear complicated to offer the possibility of repaying a home loan with annuities of over 30 years, unless the following structural problems are taken into account:

- *We need to see a sustained reduction in interest rates to a level of 2% to limit overall costs and therefore risks. A further explosion in interest rates would further contract demand for home loans, preventing the development of very long-term supply.*

- *Banks will need to be able to stabilize their capital base, as the EU tends to advise, in order to have sufficient lending capacity to avoid exacerbating the distortion between supply and demand.*

- *Provide a regulatory framework for intergenerational lending, so as not to create legal uncertainty for borrowers with no descendants, or descendants who do not wish to reclaim the property. An agreement with the banks could be included to allow resale of the property without the need to take over the loan.*

- *Even if outside the scope of this article, the opening up of bullet loans to a larger number of borrowers will need to be considered politically, given their advantages for both consumers and lenders.*

**Annexe 1 : Table relative à la figure 6.**

| Duration | Monthly_Payment | Total_Debt | Relative_Increase | Debt_Ratio | Repayment_Capacity | Risk_Index |
|---|---|---|---|---|---|---|
| 20 | 801.8224 | 192437.4 | 0.000000 | 0.1978343 | 5.054736 | 1.396476 |
| 21 | 776.6598 | 195718.3 | 1.704924 | 0.1916259 | 5.218501 | 1.866263 |
| 22 | 753.9045 | 199030.8 | 3.426274 | 0.1860115 | 5.376012 | 2.338741 |
| 23 | 733.2418 | 202374.7 | 5.163944 | 0.1809133 | 5.527508 | 2.813925 |
| 24 | 714.4092 | 205749.8 | 6.917825 | 0.1762668 | 5.673219 | 3.291828 |
| 25 | 697.1865 | 209155.9 | 8.687804 | 0.1720174 | 5.813366 | 3.772463 |
| 26 | 681.3872 | 212592.8 | 10.473763 | 0.1681192 | 5.948160 | 4.255844 |
| 27 | 666.8524 | 216060.2 | 12.275582 | 0.1645330 | 6.077807 | 4.741981 |
| 28 | 653.4459 | 219557.8 | 14.093137 | 0.1612252 | 6.202503 | 5.230883 |
| 29 | 641.0503 | 223085.5 | 15.926299 | 0.1581669 | 6.322436 | 5.722559 |
| 30 | 629.5639 | 226643.0 | 17.774937 | 0.1553328 | 6.437790 | 6.217015 |
| 31 | 618.8978 | 230230.0 | 19.638917 | 0.1527012 | 6.548739 | 6.714256 |
| 32 | 608.9746 | 233846.2 | 21.518101 | 0.1502528 | 6.655450 | 7.214284 |
| 33 | 599.7259 | 237491.5 | 23.412349 | 0.1479709 | 6.758087 | 7.717102 |
| 34 | 591.0917 | 241165.4 | 25.321517 | 0.1458405 | 6.856804 | 8.222707 |
| 35 | 583.0186 | 244867.8 | 27.245459 | 0.1438487 | 6.951751 | 8.731098 |
| 36 | 575.4591 | 248598.3 | 29.184026 | 0.1419835 | 7.043072 | 9.242270 |
| 37 | 568.3710 | 252356.7 | 31.137068 | 0.1402346 | 7.130906 | 9.756219 |
| 38 | 561.7164 | 256142.7 | 33.104431 | 0.1385927 | 7.215385 | 10.272936 |
| 39 | 555.4612 | 259955.9 | 35.085958 | 0.1370494 | 7.296639 | 10.792412 |
| 40 | 549.5750 | 263796.0 | 37.081493 | 0.1355971 | 7.374789 | 11.314636 |
| 41 | 544.0301 | 267662.8 | 39.090875 | 0.1342290 | 7.449955 | 11.839598 |
| 42 | 538.8015 | 271556.0 | 41.113944 | 0.1329389 | 7.522251 | 12.367283 |
| 43 | 533.8665 | 275475.1 | 43.150536 | 0.1317213 | 7.591785 | 12.897677 |
| 44 | 529.2045 | 279420.0 | 45.200485 | 0.1305711 | 7.658665 | 13.430764 |
| 45 | 524.7967 | 283390.2 | 47.263627 | 0.1294835 | 7.722990 | 13.966525 |
| 46 | 520.6260 | 287385.6 | 49.339793 | 0.1284545 | 7.784859 | 14.504943 |
| 47 | 516.6766 | 291405.6 | 51.428815 | 0.1274800 | 7.844365 | 15.045998 |
| 48 | 512.9342 | 295450.1 | 53.530524 | 0.1265567 | 7.901598 | 15.589670 |
| 49 | 509.3855 | 299518.7 | 55.644748 | 0.1256811 | 7.956646 | 16.135935 |
| 50 | 506.0183 | 303611.0 | 57.771316 | 0.1248503 | 8.009592 | 16.684773 |
| 51 | 502.8214 | 307726.7 | 59.910057 | 0.1240615 | 8.060516 | 17.236159 |
| 52 | 499.7845 | 311865.5 | 62.060798 | 0.1233122 | 8.109495 | 17.790068 |
| 53 | 496.8980 | 316027.1 | 64.223365 | 0.1226000 | 8.156604 | 18.346475 |
| 54 | 494.1530 | 320211.1 | 66.397585 | 0.1219228 | 8.201914 | 18.905355 |
| 55 | 491.5413 | 324417.2 | 68.583285 | 0.1212784 | 8.245493 | 19.466681 |
| 56 | 489.0552 | 328645.1 | 70.780290 | 0.1206650 | 8.287408 | 20.030424 |
| 57 | 486.6877 | 332894.4 | 72.988428 | 0.1200808 | 8.327723 | 20.596558 |
| 58 | 484.4321 | 337164.7 | 75.207524 | 0.1195243 | 8.366498 | 21.165053 |
| 59 | 482.2823 | 341455.9 | 77.437405 | 0.1189939 | 8.403792 | 21.735881 |
| 60 | 480.2325 | 345767.4 | 79.677898 | 0.1184882 | 8.439662 | 22.309012 |

**Références bibliographiques :**

1- Statista. (2024, 28 mai). *Nombre de logements vendus selon le type de marché en France 2007-2024*. https://fr.statista.com/statistiques/639701/nombre-logements-vente-france/

2- *Évolution du revenu disponible brut et du pouvoir d'achat | Insee*. (s. d.). https://www.insee.fr/fr/statistiques/2830166#:~:text=Lecture%20%3A%20au%204%E1%B5%89%20trimestre%202023,par%20rapport%20au%20trimestre%20pr%C3%A9c%C3%A9dent.

3- Igedd. (2024, 15 mai). *Prix immobilier - Évolution à long terme*. IGEDD. https://www.igedd.developpement-durable.gouv.fr/prix-immobilier-evolution-a-long-terme-a1048.html

4- *Au quatrième trimestre 2023, la baisse des prix des logements s'amplifie (-1,7 %) - Informations rapides - 73 | Insee*. (s. d.). https://www.insee.fr/fr/statistiques/7944078

5- *Chiffres clés du logement*. (2022). Le service des données et études statistiques (SDES). https://www.statistiques.developpement-durable.gouv.fr/edition-numerique/chiffres-cles-du-logement-2022/pdf/Chiffres-cles-logement-2022.pdf

6- *Publication mensuelle d'avril 2024* (s. d.). *Avril 2024* https://www.lobservatoirecreditlogement.fr/derniere-publication

7- *Panorama des prêts à l'habitat des ménages – Mars 2024 | Banque de France*. (s. d.). Banque de France. https://www.banque-france.fr/fr/publications-et-statistiques/statistiques/panorama-des-prets-lhabitat-des-menages-mars-2024

8- *La situation des grands groupes bancaires français à fin 2022*. (2022). Banque de France. https://acpr.banque-france.fr/sites/default/files/medias/documents/20230608_as147_grands_groupes_bancaires_fr_2022.pdf

9- *Crédit immobilier*. (s. d.). Direction Générale de la Concurrence, de la Consommation et de la Répression des Fraudes. https://www.economie.gouv.fr/dgccrf/Publications/Vie-pratique/Fiches-pratiques/credit-immobilier

*10-Insee - Tableau de bord de l'économie française*. (s. d.). https://www.insee.fr/fr/outil-interactif/5367857/tableau/30_RPC/31_RNP#:~:text=En%202021%2C%20en%20France%20m%C3%A9tropolitaine,de%20moins%20de%2014%20ans.

11- Simon, A. & Essafi, Y. (2017). Concurrence générationnelle et prix immobiliers. *Revue d'Économie Régionale & Urbaine*, , 109-140. https://doi.org/10.3917/reru.171.0109

12- Grjebine, T. (2015). Faut-il encourager les booms immobiliers pour sortir de la "stagnation séculaire" ?. *L'Économie politique*, 65, 7-22. https://doi.org/10.3917/leco.065.0007

13- *Le risque de crédit - 4e éd.* (s. d.). Google Books. https://books.google.fr/books?hl=fr&lr=&id=qnIhE-QET-cC&oi=fnd&pg=PR5&dq=Aghion+pr%C3%AAts+immobiliers&ots=WfXCtkOG0V&sig=Yh16NOLO-DeOjF1zPxoqYgaZD98&redir_esc=y#v=onepage&q&f=false

14- 1. Principaux éclairages sur l'action publique. (2021). *Études Économiques de L'OCDE*, *2021/12*. https://www.cairn.info/revue-etudes-economiques-de-l-ocde-2021-12-page-16.htm

15- Richesse immobilière et encours des prêts immobiliers : quelles divergences territoriales aujourd'hui ? (2018). *Revue D'économie Financière*, *2018/4*, 233-248. https://www.cairn.info/revue-d-economie-financiere-2018-4-page-233.htm

16- L'économétrie des données individuelles : l'exemple des remboursements anticipés. (1993). *Journal de la Société Statistique de Paris*, *134*. http://www.numdam.org/item/JSFS_1993__134_1_65_0.pdf

17- Comment donner une valeur à l'immobilier ? : L'efficacité des méthodes du Crédit foncier de France au xix e siècle. (2016). *Histoire & Mesure*, *2016/1*, 43-59.

https://www.cairn.info/revue-histoire-et-mesure-2016-1-page-43.htm